\documentclass[prd,twocolumn,showpacs,superscriptaddress,nofootinbib,floatfix,showkeys,10pt]{revtex4-2}
\usepackage{graphicx}
\usepackage{amsmath}
\usepackage{bm}
\usepackage{yhmath}
\usepackage{mathtools}
\usepackage{wasysym}
\usepackage[colorlinks,citecolor=violet,urlcolor=violet,linkcolor=blue]{hyperref}
\usepackage{color}
\usepackage{cases}
\usepackage{subfigure}
\usepackage{times}
\usepackage{dcolumn,booktabs,bm}
\usepackage{slashed}
\usepackage{amsfonts,amssymb,stmaryrd,latexsym,amsmath}
\usepackage{textcomp}
\usepackage{multirow}
\usepackage{cancel}
\usepackage{array}
\usepackage{orcidlink}
\usepackage{physics}
\usepackage{pstricks}
\usepackage{color}
\usepackage{amssymb}
\usepackage{epsfig}
\usepackage{mathtools}
\RequirePackage{graphicx}
\RequirePackage{mathptmx}

\begin{document}

\title{Measure of the kaon structure: generalized valence quark distribution functions and form factors}
	
	\author{H. Nematollahi\,\orcidlink{0000-0003-0162-3085}}\email{hnematollahi@uk.ac.ir}
	\thanks{Corresponding author}     
	\affiliation{Department of Physics, Shahid Bahonar University of Kerman, Kerman 76169133, Iran }

	\author{K. Azizi\,\orcidlink{0000-0003-3741-2167}}\email{kazem.azizi@ut.ac.ir}
	\thanks{Corresponding author}
		\affiliation{Department of Physics, University of Tehran, North Karegar Avenue, Tehran 14395-547, Iran}
		\affiliation{Department of Physics, Dogus University, Dudullu-\"Umraniye, 34775 Istanbul, 
	             T\"urkiye}

\begin{abstract}

We investigate the structure of the $K^+$ meson through calculating its valence-quark (anti quark) generalized parton distribution functions (GPDs) and form factors (FFs). For this purpose, we use a theoretical framework which is based on the exponential representation scheme (ERS). In this scheme the valence-quark GPDs of the light mesons, including the kaon, are considered as the valence quark distribution function crossed by the exponential of a profile function, at zero skewness. We apply the modified chiral quark model ($\chi QM$ ) to obtain the input valence distribution functions of the kaon in above scheme and calculate its valence GPDs. The kaon's electromagnetic and gravitational form factors are also determined and compared with existing experimental data and the results of some theoretical and phenomenological models.   

\end{abstract}
\maketitle
	
	\thispagestyle{empty}

\section{\label{sec-intro}Introduction}
The hadron's parton distribution functions (PDFs), which are dependent on the longitudinal momentum $x$, have been studied, extensively, in the past decades. These studies have given us a complete one-dimensional picture of hadrons. In order to achieving three-dimensional (3D) description of hadrons, the generalized parton distribution functions \cite{GPD1,GPD2,GPD3,GPD4,GPD5} have been established over two decades ago. GPDs, that are the functions of $x$, the skewness parameter $\xi$ and the square of total momentum transfer $t$, provide us the crucial information about the combinations of the transverse position and longitudinal momentum of the partons inside the hadrons \cite{LT1,LT2,LT3,LT4}. They can be accessed through exclusive processes including deeply virtual Compton scattering (DVCS) and deeply virtual meson production (DVMP), experimentally. Since experimental extraction of the GPDs is difficult due to their complicated kinematic dependence, theoretical model predictions can be useful to developing the valuable insights about GPDs.

On the other hand, different form factors of hadrons like electromagnetic form factors (EMFFs) and gravitational form factors (GFFs) are obtained from the zeroth and first-order Mellin moments of the GPDs \cite{GPD1,GPD3,GPD4,Bakulev2000,Goeke2001,Diehl2003,Belitsky2005,Boffi2007}. Hence, the GPDs are related to the energy-momentum tensor and therefore several basic quantities including spin, mass, energy and pressure because of their connection with hadron's GFFs \cite{Ji1997,BEG2018}.

Among all hadrons, probing the structure of the light pseudo-scalar mesons, including pion and kaon, has been of the great interest due to their responsibility as the Goldstone (GS) bosons, for dynamical chiral symmetry breaking. These simple bound states of a constituent quark and a constituent anti quark play an important role in the interpretation of the internal structure of the nucleons and nuclei. Therefore, the investigation of the GPDs of pion and kaon can be so useful.

There are some theoretical models, like Nambu-Jona-Lasinio (NJL) model \cite{NJL1,NJL2,NJL3} and Lattice QCD \cite{LQCD1,LQCD2} applied to calculate the GPDs and form factors of pion and kaon and some theoretical analyses that give the useful understanding of the structure of these light mesons considering their GPDs \cite{PiGPD1,PiGPD2,PiGPD3,PiGPD4,PiGPD5,PiGPD6,PiGPD7,PiGPD8,PiGPD9,PiGPD10,PiGPD11,PiGPD12,PiGPD13,
PiGPD14,PiChi1,PiChi2,PiFF1,PiFF2,PiFF3,PiFF4,Raya2024,Raya106,Raya2025}. In one of the important approaches used to achieve the GPDs of hadrons, the GPDs are determined via overlap representation in terms of light-front wave functions (LFWFs) \cite{Diehl2003,Diehl2001,LFWFs1,LFWFs2,LFWFs3,LFWFs4,LFWFs5,CMMR,CMMR1}.

We have studied the valence GPDs and FFs of the pion in our previous work \cite{NA2025} and we should note that those of nucleon have been investigated in Refs. \cite{GA22,GA231,GA232,GA24}. In present study, we compute the unpolarized valence GPDs of the kaon using a theoretical approach, called exponential representation scheme (ERS), in which the valence quark distributions of the kaon are needed as the input of the model \cite{Raya2025}. We have already obtained these valence distributions employing the modified chiral quark model \cite{NYVPi}. In this low-energy model, the structure of the kaon is considered to be composed of a bare up quark and a bare strange anti quark surrounded by clouds of Goldstone bosons \cite{Chi5,Chi6} and gluon \cite{NYVPi,NYSGPi}. The parton distribution functions of the kaon are obtained by regarding the interactions that take place in this bounded system.

The organization of this paper is as follows: The modified $\chi QM$ is reviewed in Section \ref{sec-chiQM} and the valence distributions of the kaon are obtained. In Section \ref{sec-GPDs}, we determine the valence GPDs of the kaon applying the modified $\chi QM$ in the ERS framework. In Section \ref{sec-Result}, we give the obtained results of kaon's valence GPDs. We also calculate EMFF and GFF of the kaon in this section and present the corresponding results. The conclusions is given in Section \ref{sec-con}.
\section{\label{sec-chiQM}The Valence Quark (anti quark) Distribution of Kaon}
In this section we review the approach used to obtain the valence quark distributions of the kaon in the modified chiral quark model framework \cite{NYVPi}. In this model the bare quark (\(u_0\)) and anti quark (\(\bar{s}_0\)) of the \(K^+\) meson are dressed by the clouds of GS bosons \cite{Chi5,Chi6} and gluon \cite{NYVPi,NYSGPi}. The PDFs of kaon are extracted by considering the interactions which occur in this bound system of constituent quark and anti quark and formulating them. The contributions of the fluctuation of the bare quark (anti quark) into GS boson and also the quark-anti quark pair production of the GS boson to the quark (anti-quark) distribution of the kaon are given by \cite{NYVPi,Chi5,Chi6,NYSGPi,Chi1,Chi2,Chi3,Chi4}:
\begin{equation}
q_{j}({\bar{q}}_j)(x)=P_{j{\cal
B}/i} \otimes q_{0}({\bar{q}}_0)=\int^{1}_{x}\frac{\textmd{d}y}{y}~P_{j{\cal
B}/i}(y)~q_{0,i}({\bar{q}}_{0,i})(\frac{x}{y}),\label{q}
\end{equation}
\begin{eqnarray}
q_{k}({\bar{q}}_k)(x)&=&V\otimes P\otimes
q_0({\bar{q}}_0)\nonumber\\
&=&\int\frac{\textmd{d}y_{1}}{y_{1}}\frac{\textmd{d}
y_{2}}{y_{2}}~V_{k/{\cal B}}(\frac{x}{y_{1}})~P_{{\cal B}
j/i}(\frac{y_{1}}{y_{2}})~q_{0,i}({\bar{q}}_{0,i})(y_{2}),\label{two-split}\nonumber\\
\end{eqnarray}
respectively. Here $q_{0,i}({\bar{q}}_{0,i})$ is the distribution of the bare quark (anti quark) \(i\). $P_{j{\cal B}/i}$ denotes the splitting function that gives the probability of finding the quark \(q_j\) (anti quark \(\bar q_j\)) with a GS boson ${\cal B}$ and the distribution of quark \(q_k\) (anti quark \(\bar q_k\)) in the GS boson is presented by \(V_{k/\cal B}\) \cite{NYVPi,Chi5,Chi6,NYSGPi,Chi1,Chi2,Chi3,Chi4}.

On the other hand, the direct gluon dressing effect contributes to the quark (anti quark) distribution of the kaon as \cite{NYVPi,NYSGPi}
\begin{equation}
q_{j}({\bar{q}}_j)(x)=P_{jg/i} \otimes q_{0}({\bar{q}}_0)=\int^{1}_{x}\frac{\textmd{d}y}{y}~P_{jg/i}(y)~q_{0,i}({\bar{q}}_{0,i})(\frac{x}{y}),\label{q1}
\end{equation}
and the contribution of quark–anti quark pair emission by the gluon is given as \cite{NYVPi,NYSGPi}:
\begin{eqnarray}
q_{k}({\bar{q}}_k)(x)&=&{\cal P}\otimes P\otimes
q_0({\bar{q}}_0)\nonumber\\
&=&\int\frac{\textmd{d}y_{1}}{y_{1}}\frac{\textmd{d}
y_{2}}{y_{2}}~{\cal P}_{kg}(\frac{x}{y_{1}})~P_{g
j/i}(\frac{y_{1}}{y_{2}})~q_{0,i}({\bar{q}}_{0,i})(y_{2}),\label{two-splitg}\nonumber\\
\end{eqnarray}
where ${\cal P}_{kg}$ and \(P_{jg/i}\) are the splitting functions of gluon to quarks \cite{AP} and the bare quark–gluon vertex, respectively \cite{NYVPi,NYSGPi}.

The valence quark and anti quark distributions of the kaon are calculated via the following relations regarding  Eqs.(\ref{q},\ref{two-split},\ref{q1},\ref{two-splitg}) \cite{NYVPi}:
\begin{eqnarray}
u^{K}_v(x)&=&u^{K}(x)-\bar{u}^{K}(x)\nonumber\\&
=&Z^{K}_{u}u^{K}_{0}(x)+\frac{1}{2}P_{u\pi^{0}/u}\otimes u^{K}_{0}+V_{u/\pi^{+}}\otimes P_{\pi^{+}d/u}\otimes u^{K}_{0}
\nonumber\\
&&+V_{u/K^{+}}\otimes P_{K^{+}\bar{u}/\bar{s}}\otimes \bar{s}^{K}_{0}+V_{u/K^{+}}\otimes
P_{K^{+}s/u}\otimes u^{K}_{0} \nonumber
\\&&
+P_{{ug}/{u}}\otimes u^{K}_{0}-P_{\bar{u}K^{+}/\bar{s}}\otimes \bar{s}^{K}_{0}
\;,\label{uvk(x)}
\end{eqnarray}
\begin{eqnarray}
\bar{s}^{K}_v(x)&=&\bar{s}^{K}(x)-s^{K}(x)\nonumber\\&
=&Z^K_{s}\bar{s}^{K}_{0}+V_{\bar{s}/k^{0}}\otimes P_{k^{0}\bar{d}/\bar{s}}\otimes
\bar{s}^{K}_{0} +V_{\bar{s}/k^{+}}\otimes P_{k^{+}s/u}\otimes
u^{K}_{0}\nonumber\\&&+V_{\bar{s}/k^{+}}\otimes P_{k^{+}\bar{u}/\bar{s}}\otimes
\bar{s}^{K}_{0}+P_{{\bar{s}g}/{\bar{s}}}\otimes \bar{s}^{K}_{0}-P_{sK^{+}/u}\otimes u^{K}_{0}.\nonumber\\\label{svk(x)}
\end{eqnarray}
\(Z_u^K\) and \(Z_s^K\) are the renormalization factors of the bare \(u\) quark and \(\bar s\) anti quark which are computed from the following relation that is satisfied by the valence distributions of \(K^+\) \cite{Chi6,NYVPi,HCT,Chenetal}:
\begin{equation}
\int^{1}_{0} u^{K}_v(x) \textmd{d}x=\int^{1}_{0} \bar{s}^{K}_v(x) \textmd{d}x=1.\label{vanorcon}
\end{equation}

\section{\label{sec-GPDs}Valence-quark GPDs of Kaon}
In this section the valence-quark (anti quark) GPDs of the kaon are obtained applying the exponential representation scheme \cite{Raya2025}. Let us first start from the overlap representation of GPDs for kaon within DGLAP domain at renormalization scale $\mu$ as \cite{PiGPD8,PiGPD12,PiGPD13,Raya2025,NA2025,Diehl2003,LT3}:
\begin{eqnarray}
H_q^{K}(x,{\xi},t;\mu)=
\int \frac{d^2k_\perp}{16{\pi}^3} {{\psi}_q^{K}}^{\ast}(x_-,k^2_{{\perp}-};\mu){\psi}_q^{K}(x_+,k^2_{{\perp}+};\mu),\nonumber\\\label{LFWF}\end{eqnarray}
where the LFWF is denoted by $\psi$, $t=-\Delta^{2}$ and $\xi$ are the momentum transfer squared and the skewness parameter, respectively, and $x_{\pm}=\frac{x\pm\xi}{1\pm\xi}$, $k_{{\perp}\pm}=k_{\perp}\mp\frac{1-x_{\pm}}{1\pm\xi}\frac{\Delta_{\perp}}{2}$.

Then above LFWF is written in terms of the valence quark distribution function of the kaon $q^{K}(x;\mu)$ and a profile function $\hat{\phi}^{K}(x;\mu)$ within the ERS as \cite{Raya2025}:
\begin{eqnarray}
{\psi}_q^{K}(x,k^2_{{\perp}};\mu)=8 \pi \frac{\sqrt{q^{K}(x;\mu) \hat{\phi}^{K}(x;\mu)}}{1-x}
\exp[-2k^2_{{\perp}}\frac{\hat{\phi}^{K}(x;\mu)}{(1-x)^2}].\nonumber\\ \label{LFWF1}
\end{eqnarray}
We should emphasize that the relation between the leading-twist kaon distribution amplitude (DA) $\varphi_{q}^{K}(x;\mu)$ and the LFWF is given by \cite{PiGPD8,PiGPD13,Raya2025,Brodsky1989}:
\begin{equation}
f^K \varphi_{q}^{K}(x;\mu)=\int \frac{d^2k_\perp}{16{\pi}^3}{\psi}_q^{K}(x,k^2_{{\perp}};\mu), \label{DA}
\end{equation} 
in which $f^{K}$ is the leptonic decay constant of the kaon. In the ERS framework the distribution amplitude of the kaon is given in terms of its valence distribution as \cite{Raya2025}:
\begin{equation}
\varphi_{q}^{K}(x;\mu)=\frac{1}{4\pi f^K}(1-x)\sqrt{\frac{q^{K}(x;\mu)}{\hat{\phi}^{K}(x;\mu)}}. \label{DA1}
\end{equation} 
So by choosing the profile function $\hat{\phi}^{K}$, the DA is completely determined.

By regarding Eqs.(\ref{LFWF},\ref{LFWF1}), the valence GPDs of the kaon at zero skewness is obtained as \cite{Raya2025}:
\begin{equation}
H_q^{K}(x,t;\mu)=q^{K}(x;\mu)\exp[t\hat{\phi}^{K}(x;\mu)]. \label{KaGPD}
\end{equation}
It should be noted that in the forward limit (\(t=0\)), $H_q^K(x,0;\mu)=q^K(x;\mu)$ as expected and the exponential term in the above equation controls the evolution of the valence GPDs.
	
In Ref.\cite{Raya2025}, the profile function for light mesons, like the kaon, is suggested as $\hat{\phi}^{K}(x;\mu)=(1-x)^2/\Lambda_K^2$, in which $\Lambda_K$ is the mass scale. Hence, the valence GPDs of the kaon is obtained as \cite{Raya2025}
\begin{equation}
H_q^{K}(x,t;\mu)=q^{K}(x;\mu)\exp[\frac{t}{\Lambda_K^2}(1-x)^2]. \label{KaGPDM}
\end{equation}	
Considering above equation, by having the valence quark (anti quark) distribution functions \(q^K(x;\mu)\), we can calculate the valence GPDs of kaon. These valence distributions are first obtained applying the approach described in the last section and then the kaon's valence GPDs are calculated employing Eq.(\ref{KaGPDM}).
\begin{figure*}[htp]
  \begin{center}
    \begin{tabular}{cc}
      {\includegraphics[width=70mm,height=70mm]{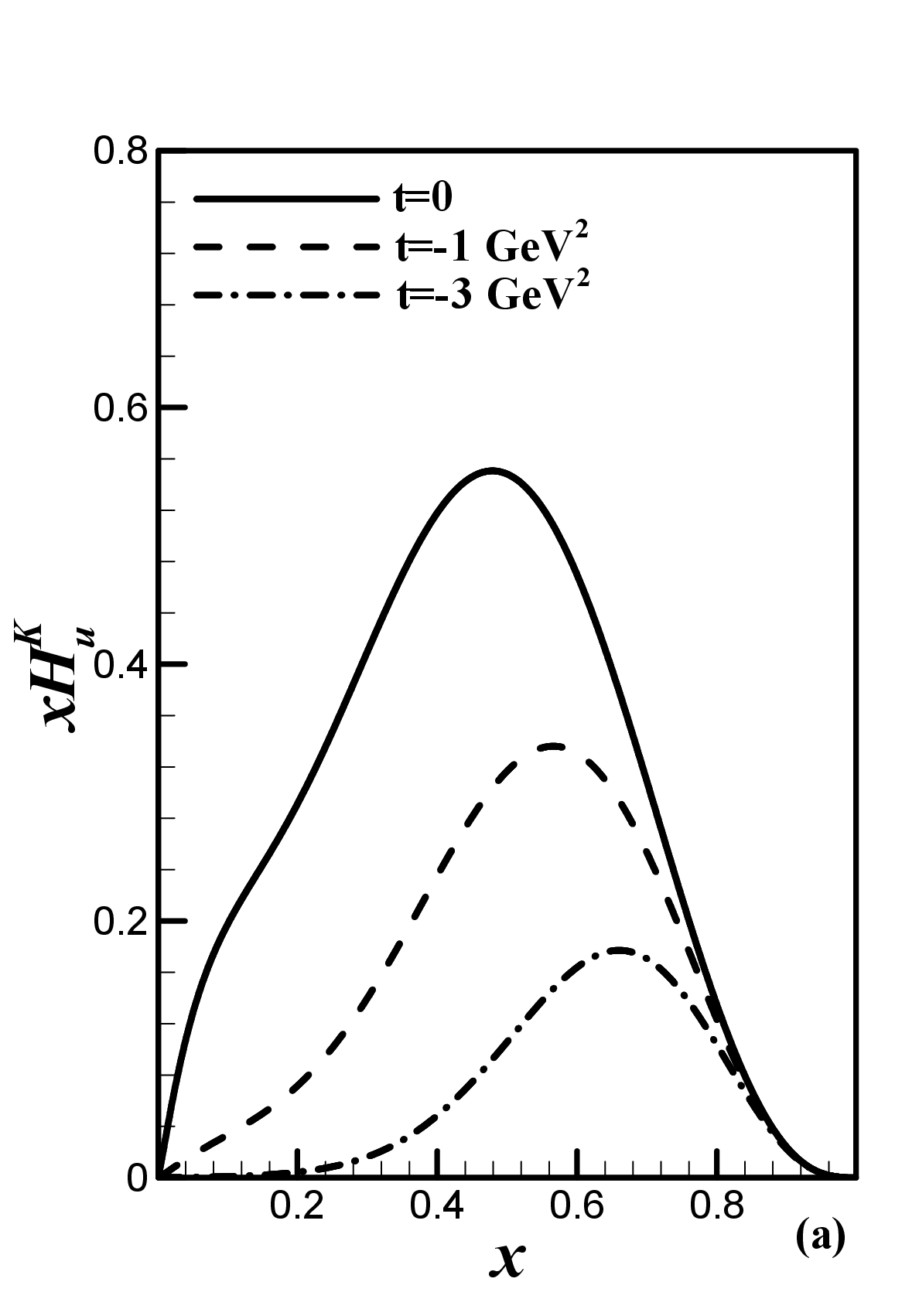}} &
       {\includegraphics[width=70mm,height=70mm]{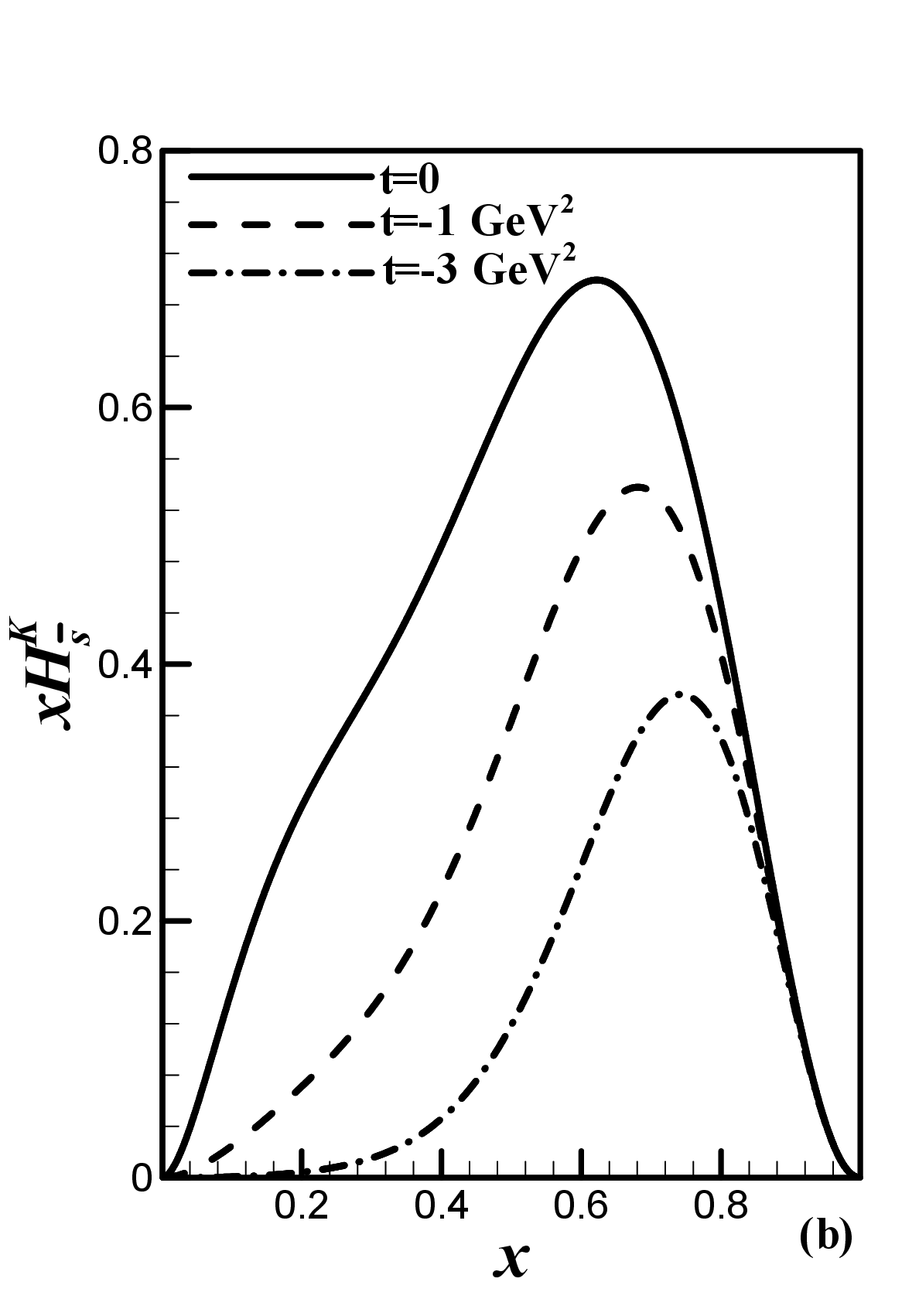}} \\
     \end{tabular}
\caption{ The results of our work for the $x$ dependence of the valence GPDs of the kaon at fix $t$ values: (a) $xH_u^{K}$ and (b) $xH_{\bar{s}}^{K}$, at $\mu_0^{2}=0.25~\textrm{GeV}^2$.  \label{fig:1}}
      \end{center}
\end{figure*}

\begin{figure*}[htp]
  \begin{center}
    \begin{tabular}{cc}
      {\includegraphics[width=70mm,height=70mm]{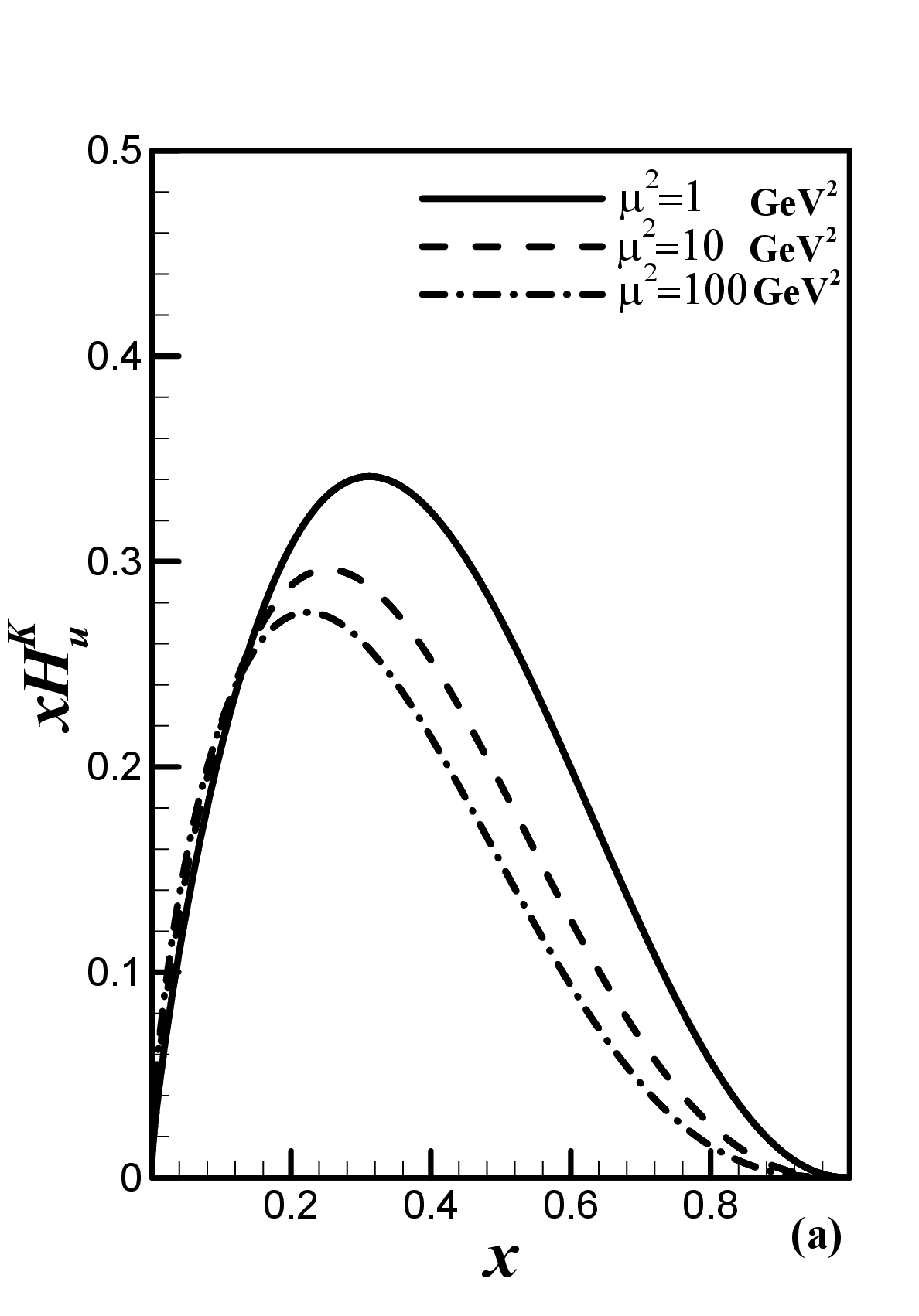}} &
       {\includegraphics[width=70mm,height=70mm]{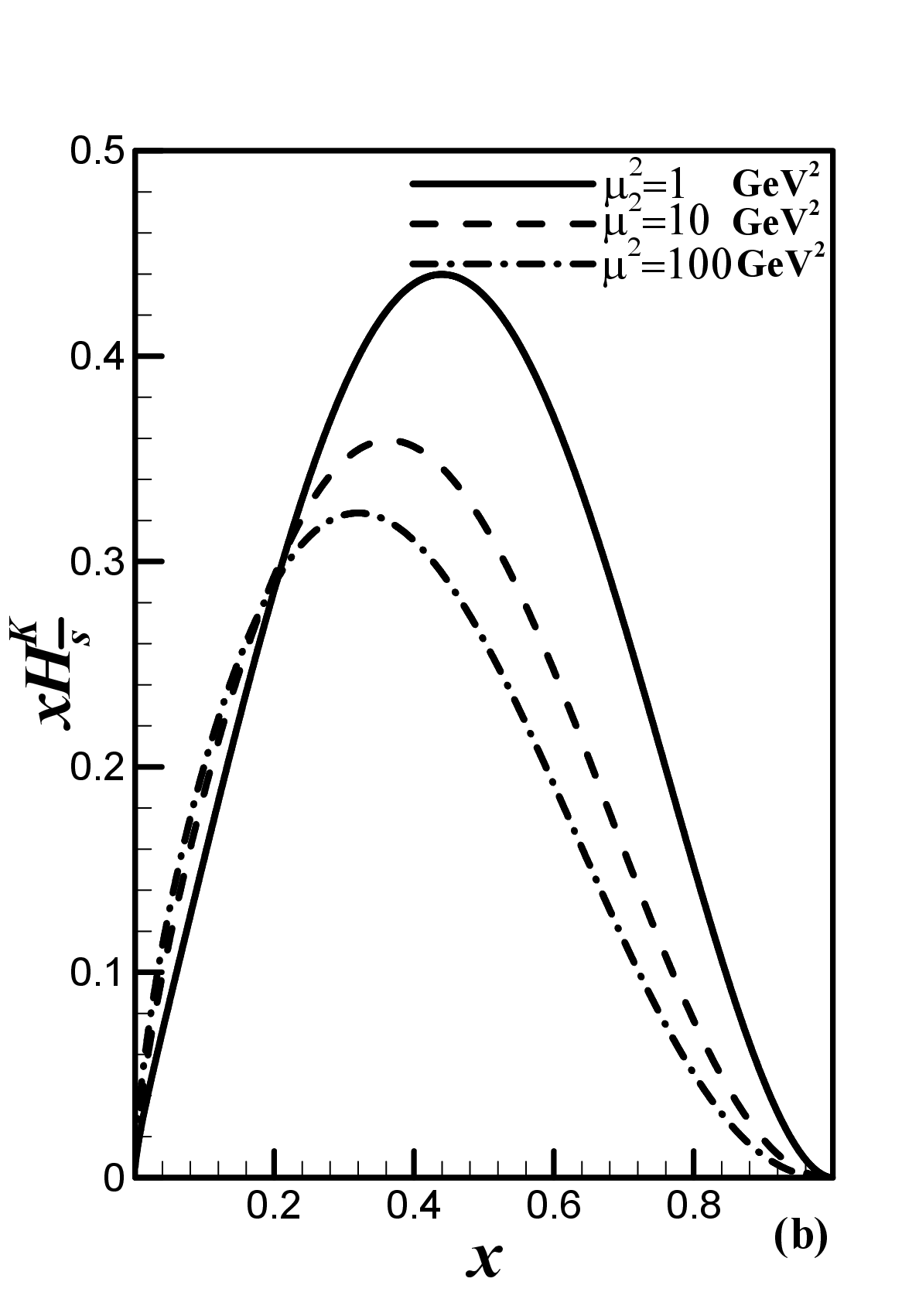}} \\
     \end{tabular}
\caption{ The evolved valence GPDs of the kaon at three $\mu^2$ values: (a) $xH_u^{K}$ and (b) $xH_{\bar{s}}^{K}$, at $t=-0.11~\textrm{GeV}^2$.  \label{fig:2}}
      \end{center}
\end{figure*}

\begin{figure*}[htp]
  \begin{center}
    \begin{tabular}{ccc}
      {\includegraphics[width=57mm,height=65mm]{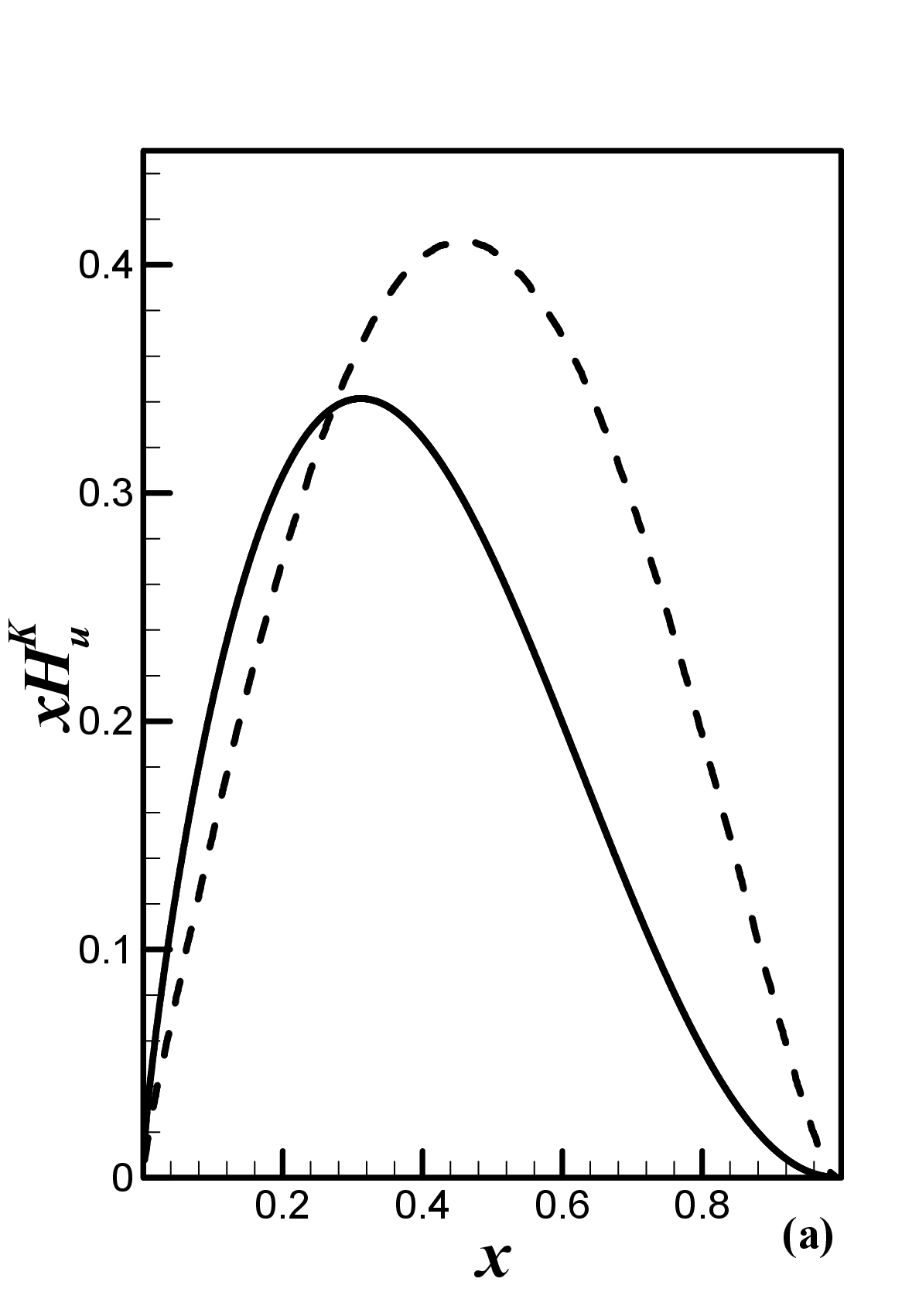}} &
      {\includegraphics[width=57mm,height=65mm]{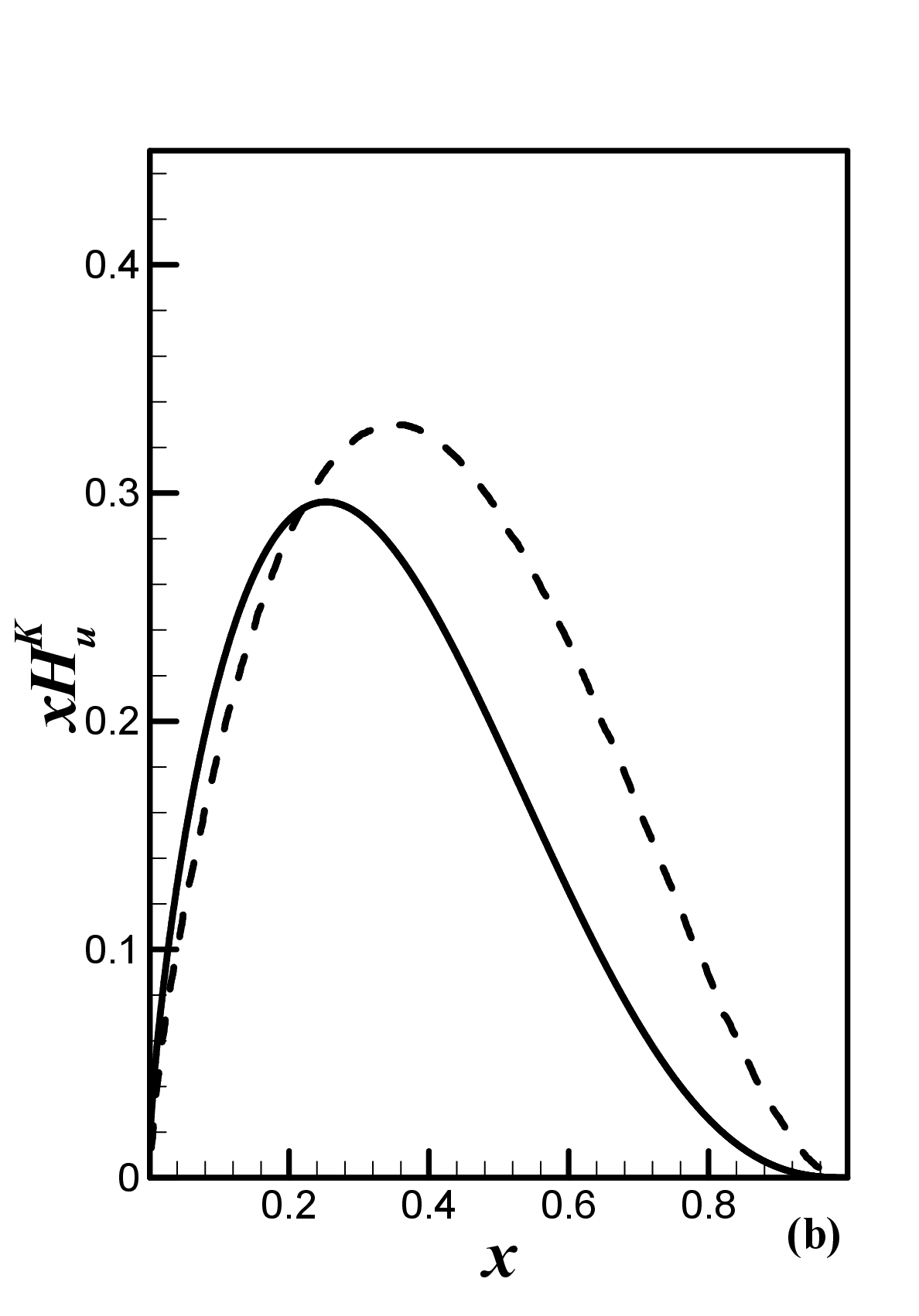}} &
       {\includegraphics[width=57mm,height=65mm]{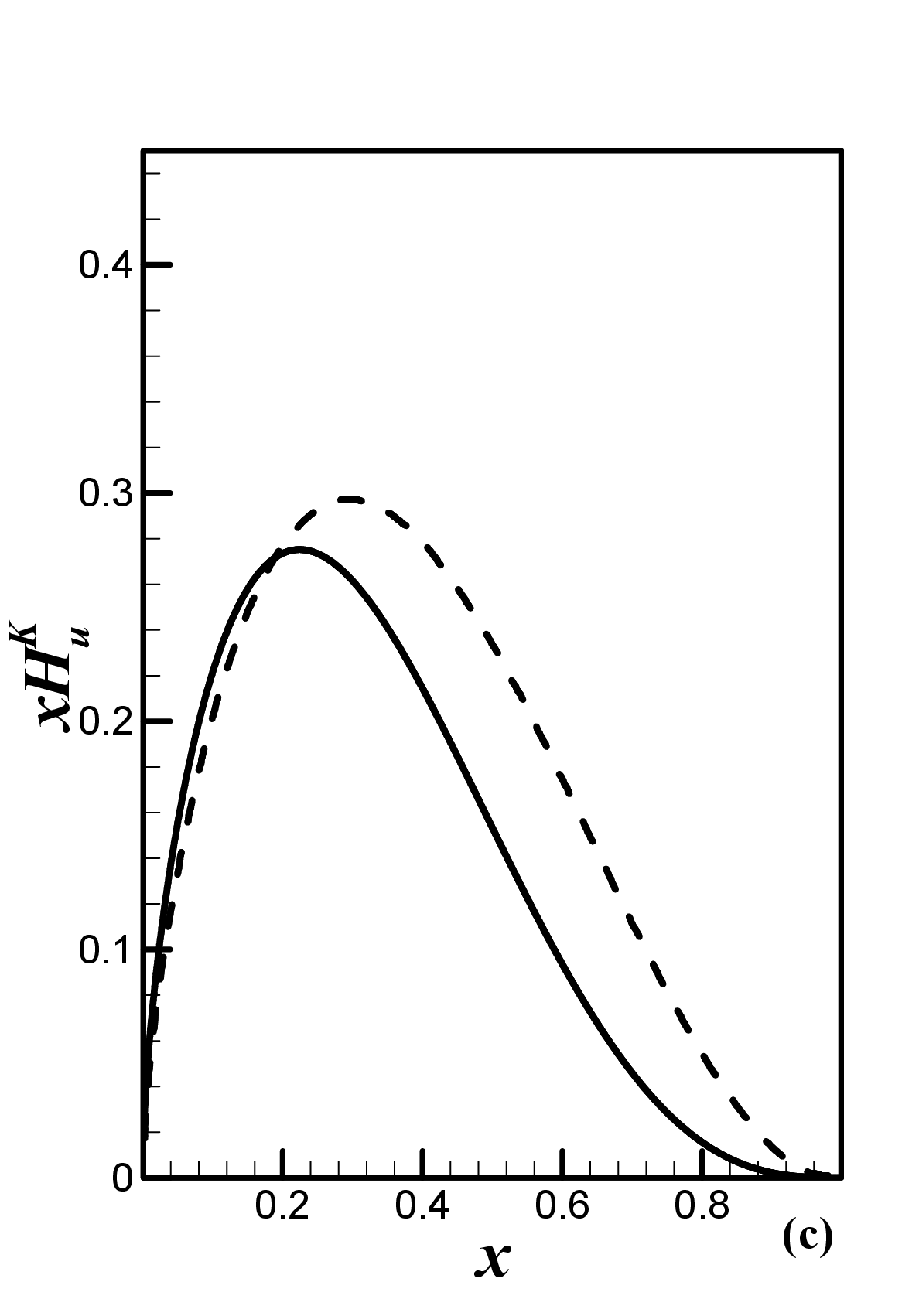}} \\
     \end{tabular}
\caption{ $xH_u^{K}$ in terms of $x$ at (a) $\mu^{2}=1~\textrm{GeV}^2$, (b) $\mu^{2}=10~\textrm{GeV}^2$ and (c) $\mu^{2}=100~\textrm{GeV}^2$. Our results (solid lines) are compared with those of Ref. \cite{PiGPD11} (dashed lines), at $t=-0.11~\textrm{GeV}^2$.  \label{fig:3}}
      \end{center}
 \end{figure*}
      
\begin{figure*}[htp]
  \begin{center}
    \begin{tabular}{ccc}
      {\includegraphics[width=57mm,height=65mm]{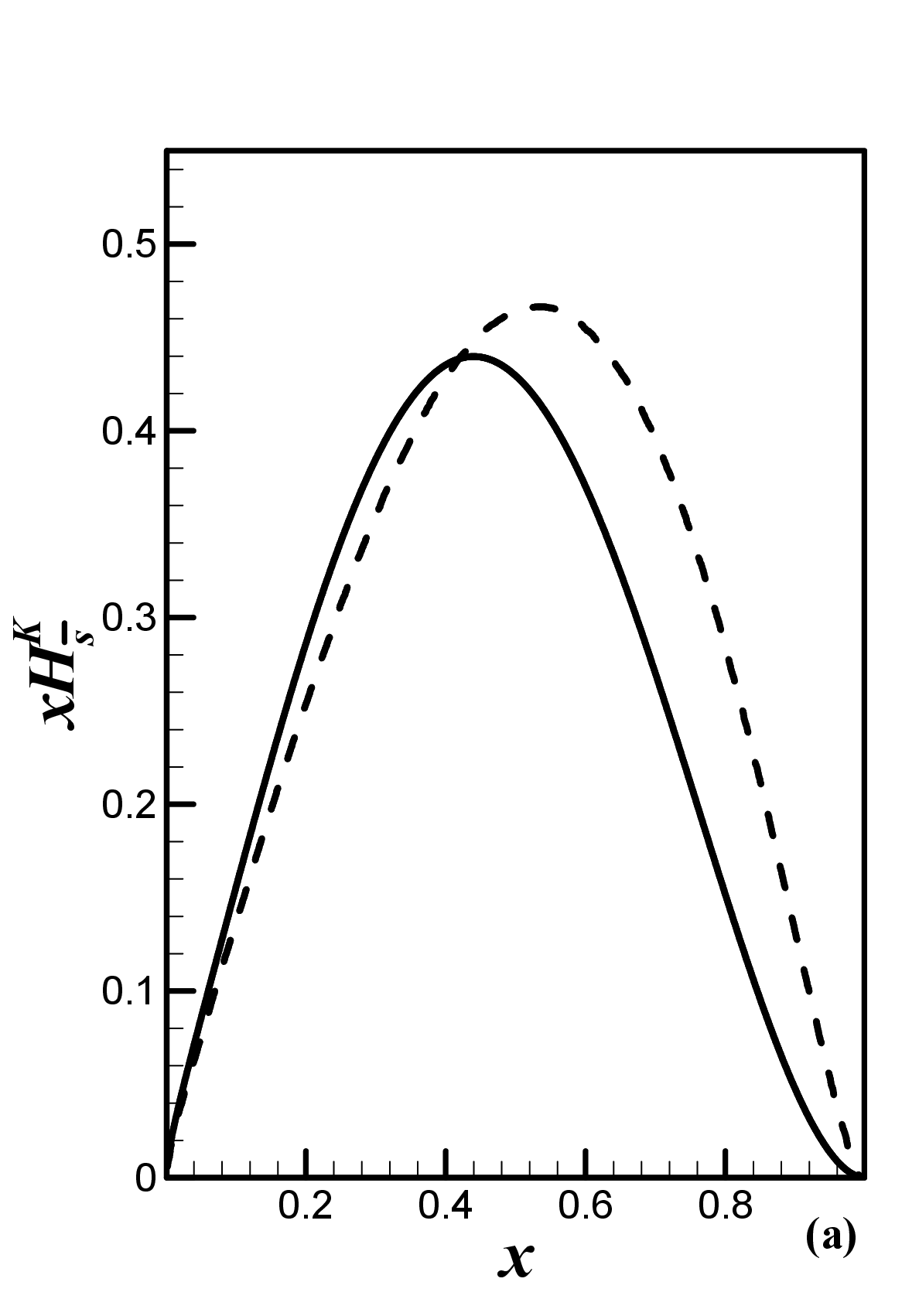}} &
      {\includegraphics[width=57mm,height=65mm]{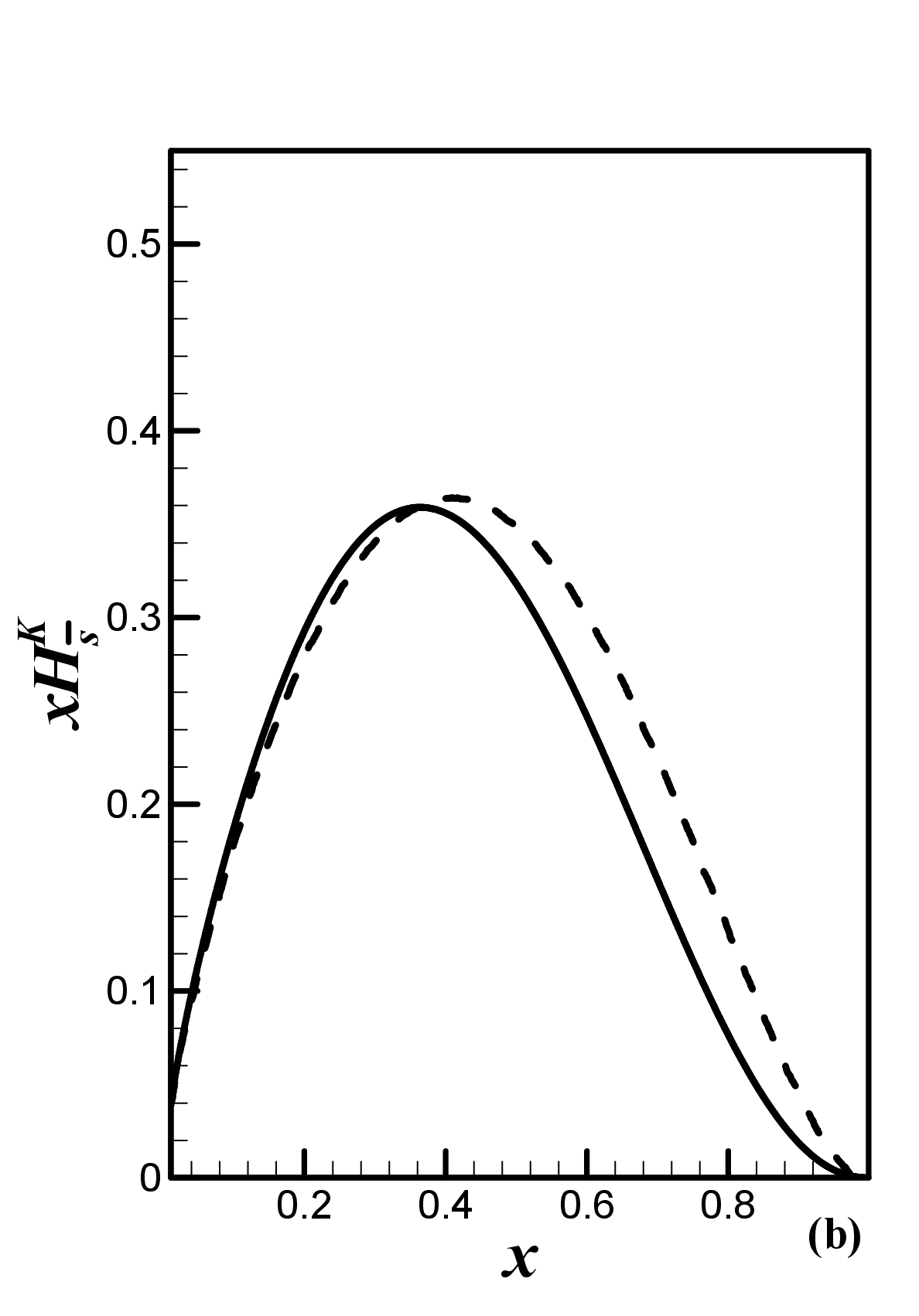}} &
       {\includegraphics[width=57mm,height=65mm]{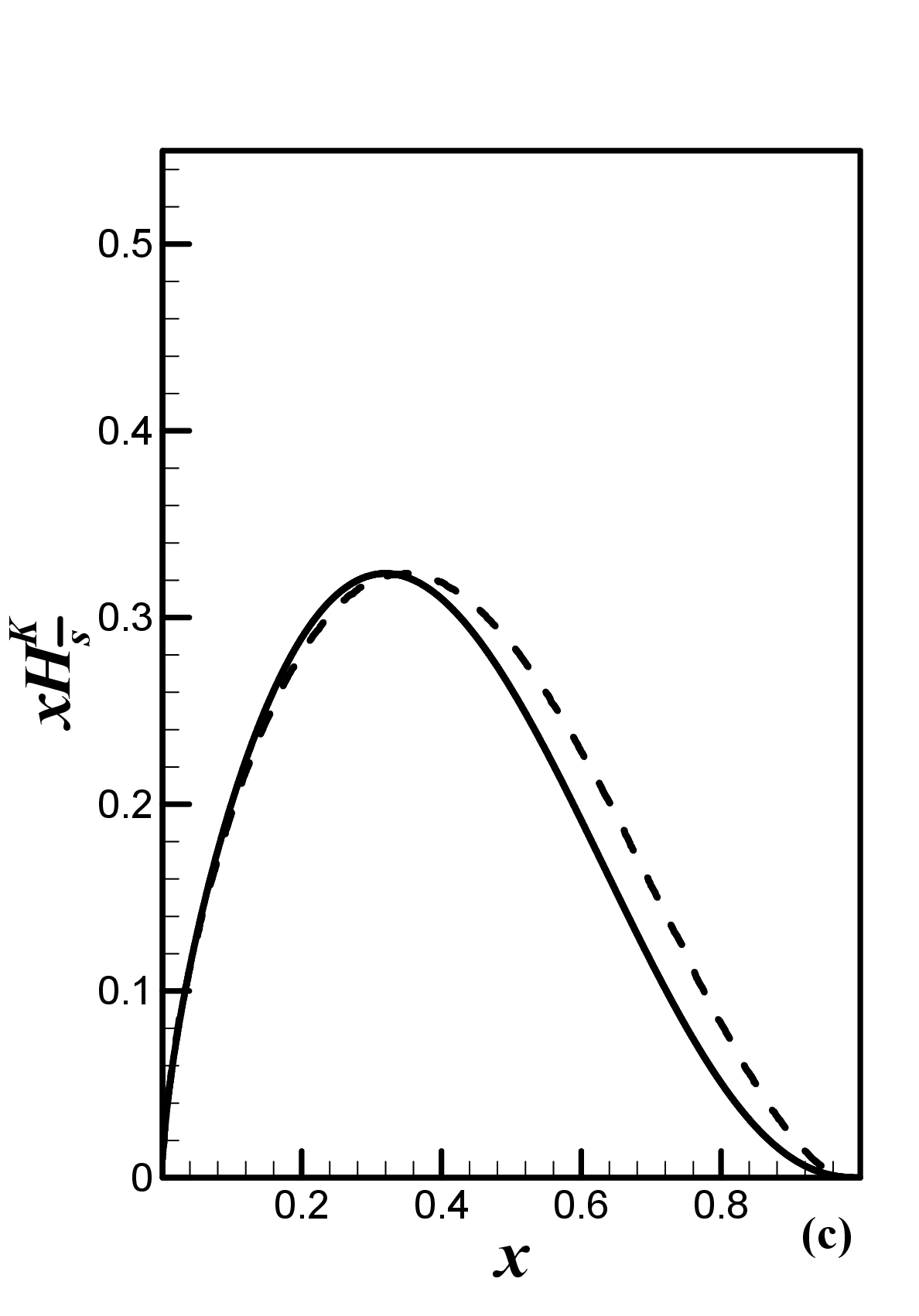}} \\
     \end{tabular}
\caption{ $xH_{\bar{s}}^{K}$ in terms of $x$ at (a) $\mu^{2}=1~\textrm{GeV}^2$, (b) $\mu^{2}=10~\textrm{GeV}^2$ and (c) $\mu^{2}=100~\textrm{GeV}^2$ scales. The results of present study (solid lines) are shown in comparison with those of Ref. \cite{PiGPD11} (dashed lines), at $t=-0.11~\textrm{GeV}^2$.  \label{fig:4}}
      \end{center}
 \end{figure*}
 
\begin{figure*}[htp]
  \begin{center}
    \begin{tabular}{c}
      {\includegraphics[width=86mm,height=70mm]{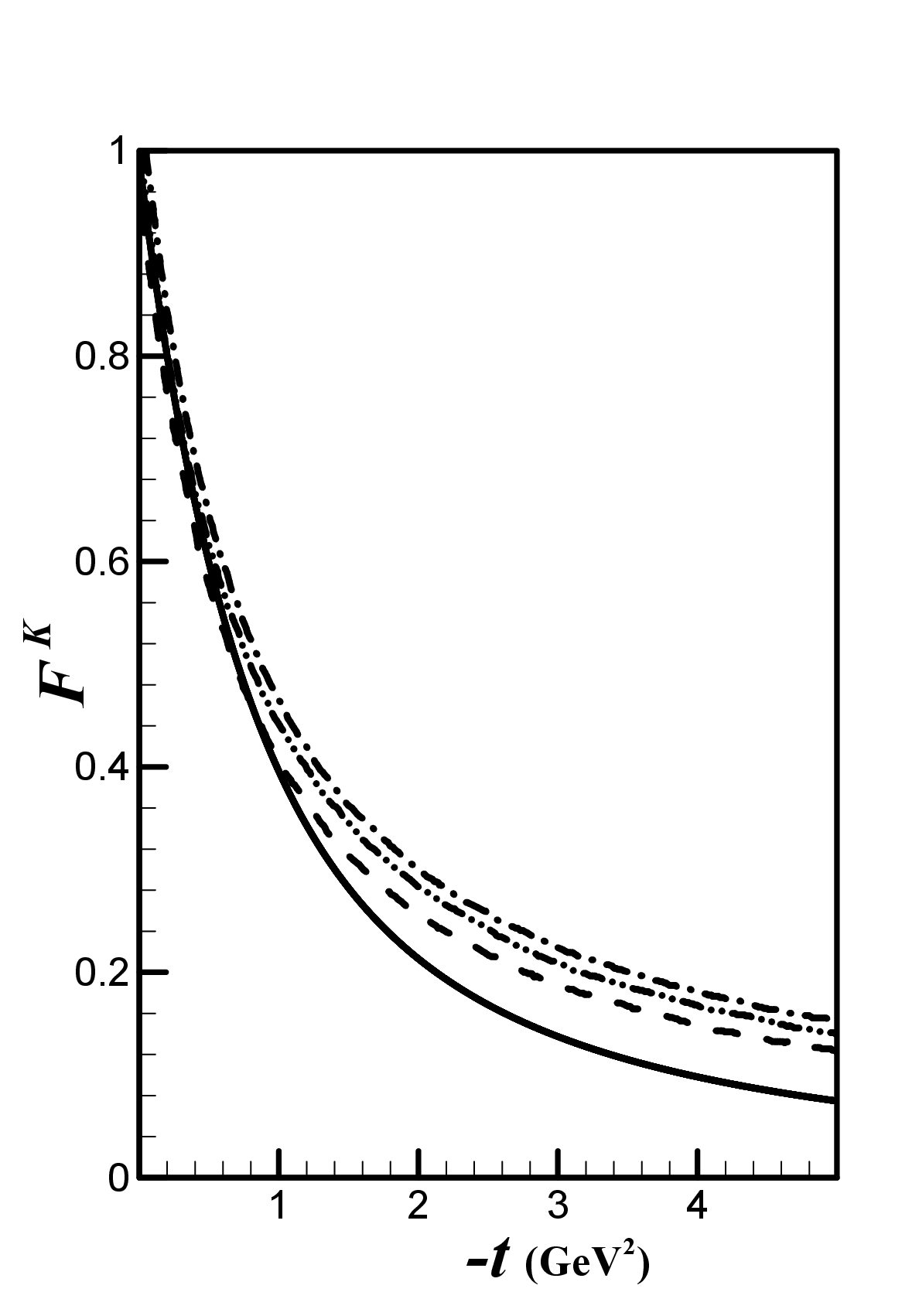}} \\
        \end{tabular}
\caption{ The EFF of kaon ($F^{K}(t)$). Our result (solid line) is compared with the NJL (dashed line) \cite{HCT} and Lattice QCD (dash-dot-dotted line) \cite{Alex2022} results and also that of Ref.\cite{ErA2022} (dash-dotted line). \label{fig:5}}
      \end{center}
\end{figure*}
\begin{figure*}[htp]
  \begin{center}
    \begin{tabular}{cc}
      {\includegraphics[width=70mm,height=70mm]{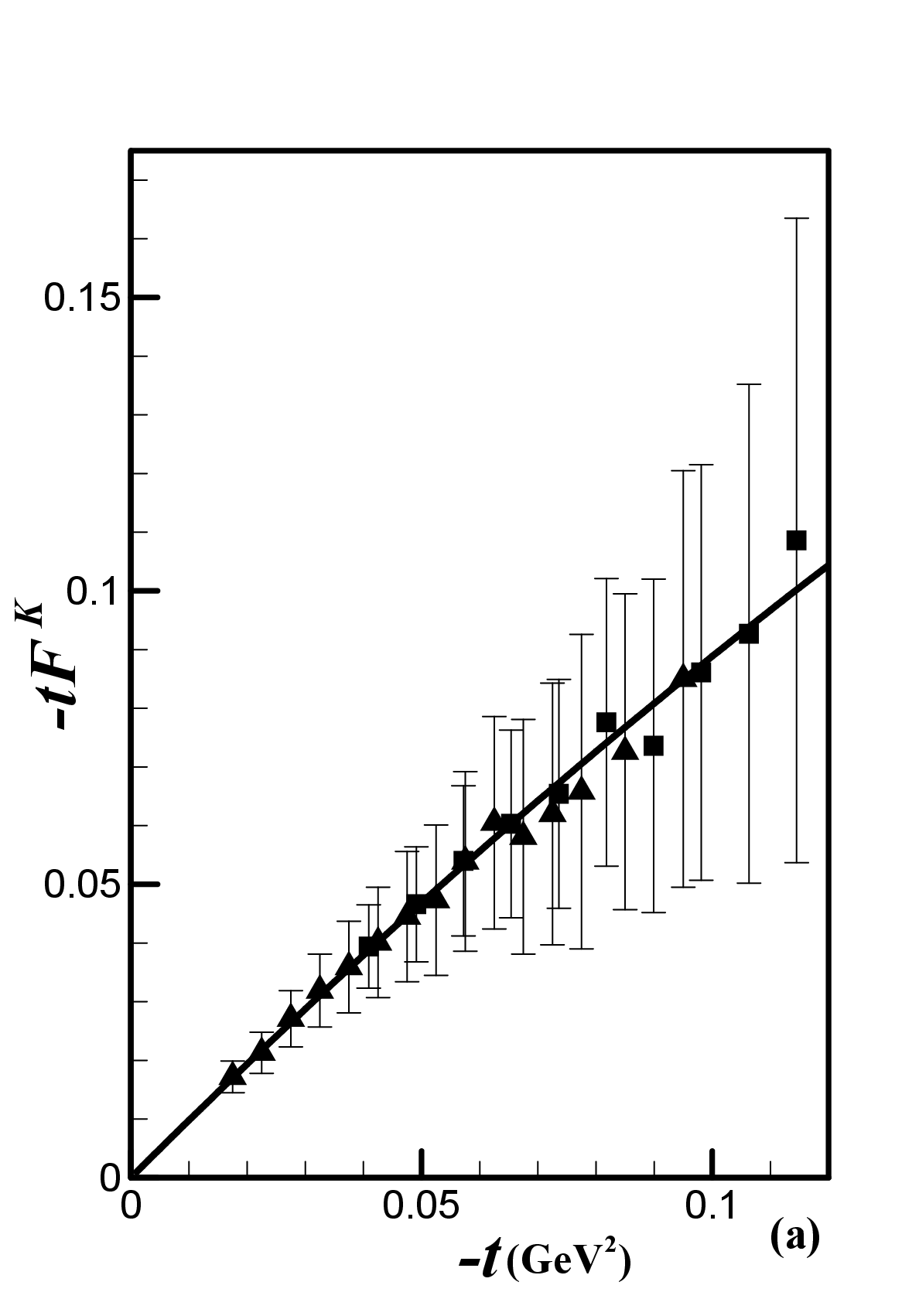}} &
       {\includegraphics[width=70mm,height=70mm]{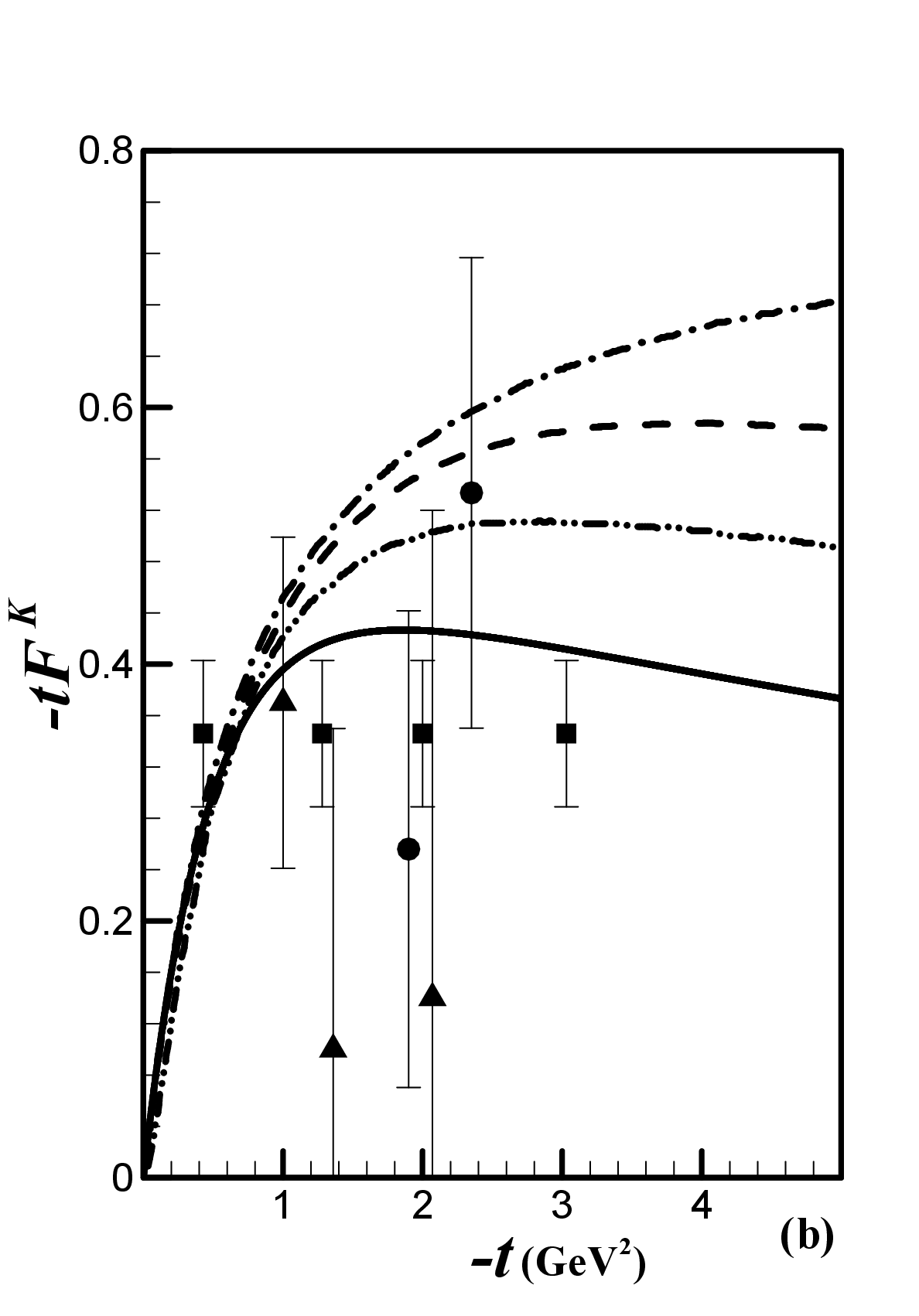}} \\
     \end{tabular}
\caption{ (a) $-tF^{K}$ with respect to $-t$ within $0\leqslant -t\leqslant0.12~\textrm{GeV}^2$. Our result (solid line) is presented in comparison with the two data sets taken from Ref.\cite{Dally1980} (filled squares) and Ref.\cite{Amend1986} (filled triangles) (b) $-tF^{K}$ as a function of $-t$ within $0\leqslant -t\leqslant5~\textrm{GeV}^2$. The result of present study (solid line) is compared with those of Ref.\cite{ErA2022} (dash-dotted line), Ref.\cite{Gao2017} (dash-dot-dotted line) and Ref.\cite{PiGPD11} (dashed line). The filled circles and triangles are corresponding to two sets of JLAB data \cite{JLAB1,JLAB2}. The filled squared represent the $t$ range and projected uncertainties of the E12-09-011 JLAB experiment \cite{JLAB1,ErA2022}. \label{fig:6}}
      \end{center}
\end{figure*}
\begin{figure*}[htp]
  \begin{center}
    \begin{tabular}{c}
      {\includegraphics[width=70mm,height=70mm]{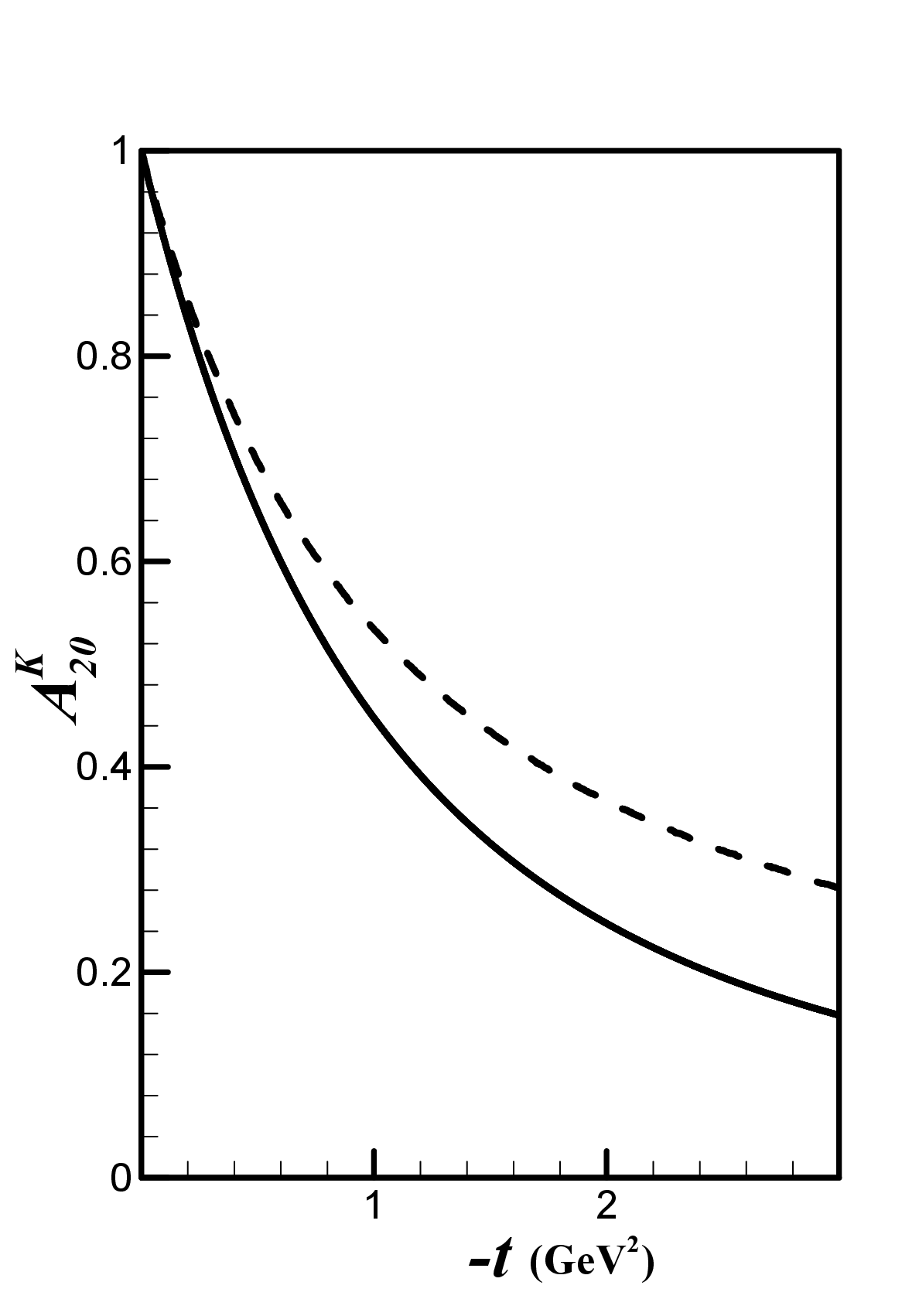}} \\
        \end{tabular}
\caption{ The complete GFF of the kaon ($A_{20}^{K}(t)$) at $\mu^{2}=4~\textrm{GeV}^2$. Our result (solid line) is compared to that of Ref.\cite{PiGPD13} (dashed line) \label{fig:7}}
      \end{center}
\end{figure*}
\section{\label{sec-Result}Results and Discussion}
We calculate the unpolarized valence-quark (anti quark) GPDs of the $K^+$ meson applying Eq.(\ref{KaGPDM}). For this purpose, we need to have the valence quark (anti quark) distribution, $u_v^{K}$ (${\bar{s}}_v^{K}$), of the kaon. We have obtained these valence distributions based on the  modified $\chi QM$ at the scale $\mu_0^{2}=0.25~\textrm{GeV}^2$ \cite{NYVPi} and therefore we use them for present calculations.

We present the results of valence GPDs of the kaon, $xH_u^{K}(x,t)$ and $xH_{\bar{s}}^{K}(x,t)$, in Figs.\ref{fig:1}a and \ref{fig:1}b, respectively. In Fig.\ref{fig:1}, these valence GPDs are shown at $\mu_0^{2}=0.25~\textrm{GeV}^2$ as the functions of $x$ at fix $t$ values: $t=0, t=-1~\textrm{GeV}^2$ and $t=-3~\textrm{GeV}^2$. As can be seen in this figure, the peaks of the valence GPDs move to the larger values of $x$ by increasing the value of $-t$. We also observe that the position of peak of $xH_u^{K}$ is located at smaller-$x$ value in comparison with that of $xH_{\bar{s}}^{K}$ at each value of $t$. As we have pointed before, the valence GPDs at $t=0$ give the valence distribution functions.

We evolve the kaon's valence GPDs from $\mu_0^{2}=0.25~\textrm{GeV}^2$ to the higher values of $\mu^2$, $\mu^{2}=1~\textrm{GeV}^2$, $\mu^{2}=10~\textrm{GeV}^2$ and $\mu^{2}=100~\textrm{GeV}^2$ using DGLAP evolution equations \cite{Chi4,AP} and present the results for $xH_u^{K}$ and $xH_{\bar{s}}^{K}$ with respect to $x$ at $t=-0.11~\textrm{GeV}^2$ in Fig.\ref{fig:2}. It is shown that for both valence-quark and anti quark GPDs, the position of peaks shift to smaller $x$ values by increasing the $\mu^{2}$ scale as expected.

In Figs.\ref{fig:3} and \ref{fig:4}, we depict the results of our work on $xH_u^{K}$ and $xH_{\bar{s}}^{K}$ at three $\mu^{2}$ scales, respectively and compare our results , at each $\mu^{2}$ value with those of Ref.\cite{PiGPD11}. From these figures, we observe that our results for $xH_u^{K}$ ($xH_{\bar{s}}^{K}$) show good agreement with the results of Ref.\cite{PiGPD11} especially within $x<0.25$ ($x<0.4$).

In the next step, we calculate the EMFF of kaon using following equation \cite{PiGPD7,PiGPD11,PiGPD12,PiGPD13,NA2025}:   
\begin{equation}
F^{K}(t)=e_u~F_u^{K}(t)+e_{\bar{s}}~F_{\bar{s}}^{K}(t),\label{EFFKa}
\end{equation}
where the charge of valence quark (anti quark) is denoted by $e_q$ ($e_{\bar{q}}$) and $F_q^{K}$ ($F_{\bar{q}}^{K}$) is the contribution of the valence quark (anti quark) to the EFF of the kaon obtained via the zeroth moment of the valence GPDs as \cite{PiGPD7,PiGPD11,PiGPD12,PiGPD13,NA2025}: 
\begin{equation}
F_{q(\bar{q})}^{K}(t)=\int_{-1}^{1} dx~H_{q(\bar{q})}^{K}(x,t).\label{EFFq}
\end{equation}
We present the result of our theoretical model on $F^{K}(t)$ in comparison with the NJL \cite{HCT} and Lattice QCD \cite{Alex2022} results and also that of Ref.\cite{ErA2022} in Fig.\ref{fig:5}. From this figure we see a good consistency among our result and presented results of other models. 

In Fig.\ref{fig:6}a the result of our work on $-tF^{K}(t)$ is depicted with respect to $-t$ at its low values (in the interval $0\leqslant -t\leqslant0.12~\textrm{GeV}^2$) and compared with the available experimental data \cite{Dally1980,Amend1986}. It is seen a good agreement between our result and the experimental data. We also present $-tF^{K}(t)$ with respect to $-t$ in the interval $0\leqslant -t\leqslant5~\textrm{GeV}^2$ and compare our result with existing data \cite{JLAB1,JLAB2} and the results of some theoretical and phenomenological models \cite{PiGPD11,ErA2022,JLAB1,Gao2017}, in Fig.\ref{fig:6}b. We see a good agreement between the result of our work and the presented experimental data. The prediction of our theoretical model show appropriate consistency with those of other models within $-t<1~\textrm{GeV}^2$. For $-t>1~\textrm{GeV}^2$ the above models give different predictions on the behavior of $-tF^{K}(t)$.

The charge radius of the kaon, $r_{K}^2=-6\frac{dF^{K}(t)}{dt}\mid_{t=0}$ \cite{PiGPD7}, is also computed via our theoretical model and $r_{K}\simeq0.577~\textrm{fm}$ is obtained. This result is in good agreement with the experimental values of $r_{K}$: $r_{K}\simeq0.56~\textrm{fm}$, $r_{K}\simeq0.53~\textrm{fm}$ \cite{PiGPD13,PiRad1,PiRad2} and $r_{K}\simeq0.58~\textrm{fm}$ \cite{Raya2025}.

Finally, we determine the complete gravitational form factor of the kaon that is obtained from computing the first-order Mellin moments of its valence quark (anti quark) GPDs as \cite{PiGPD13,Raya2025}:    
\begin{equation}
A_{20}^{K}(t)=A_{20,u}^{K}(t)+A_{20,\bar{s}}^{K}(t),\label{GFFKa}
\end{equation}
where
\begin{equation}
A_{20,q(\bar{q})}^{K}(t)=\int_{-1}^{1} dx~x~H_{q(\bar{q})}^{K}(x,t).\label{GFFq}
\end{equation}
The result of our work for $A_{20}^{K}$ with respect to $-t$ is presented in Fig.\ref{fig:7}, at the scale $\mu^{2}=4~\textrm{GeV}^2$ and compared with that of Ref.\cite{PiGPD13}. An appropriate consistency is seen among the results of our model and Ref.\cite{PiGPD13}. 

\section{\label{sec-con}Conclusion}
we have investigated the internal structure of the kaon by linking two theoretical framework: the modified chiral quark model and the exponential representation scheme, for the first time. In order to extracting the valence GPDs of kaon in the ERS framework, that relates directly the valence GPDs to the valence distribution functions, it is needed to have suitable valence distributions as the input. We have employed the results of the modified $\chi QM$ for these valence distributions of kaon at the scale $\mu_0^{2}=0.25~\textrm{GeV}^2$. The obtained DGLAP-domain valence-quark (anti quark) GPDs of the kaon via this completely theoretical model, show appropriate general properties. We have evolved these valence GPDs of kaon to the higher $\mu^2$ values using DGLAP evolution equations. The results of the kaon's evolved valence GPDs are in good agreement with the results of Ref.\cite{PiGPD11}. 

We have also extracted the EMFF and GFF of the kaon by computing the zeroth and first-order Mellin moments of its valence GPDs and compared the results with the available experimental data and the results of some other theoretical and phenomenological models. We have observed a good consistency among our results and those of other presented models.

It should be noted that there are limited theoretical models for determining the GPDs and FFs of the light mesons, especially the kaon, and therefor accessing to important information that they contain about the structure of this meson. Hence, the theoretical studies that probe the GPDs and FFs of the kaon like present work can be useful in order to giving us the better understanding of remarkable properties of these GPDs and FFs.  
%
%
\section*{Acknowledgments}
H. Nematollahi would like to thank M. M. Yazdanpanah for interesting discussion and useful comments.

\appendix


\begin{thebibliography}{92}
\bibitem{GPD1} D. M{\"u}ller et al.,
\href{https://doi.org/10.1002/prop.2190420202} {Fortsch. Phys. {\bf 42}, 101 (1994)}.
\bibitem{GPD2} A. V. Radyushkin, 
\href{https://doi.org/10.1016/0370-2693(96)00844-1} {Phys. Lett. B {\bf 385}, 333 (1996)}.
\bibitem{GPD3} A. V. Radyushkin, 
\href{https://doi.org/10.1103/PhysRevD.56.5524} {Phys. Rev. D {\bf 56}, 5524 (1997)}.
\bibitem{GPD4} X. D. Ji, 
\href{https://doi.org/10.1103/PhysRevLett.78.610} {Phys. Rev. Lett. {\bf 78}, 610 (1997)}.
\bibitem{GPD5} X. D. Ji, 
\href{https://doi.org/10.1088/0954-3899/24/7/002} {J. Phys. G {\bf 24}, 1181 (1998)}.
\bibitem{LT1} M. Burkardt, 
\href{https://doi.org/10.1103/PhysRevD.62.071503}{Phys. Rev. D {\bf 62}, 071503 (2000); {\bf 66}, 119903 (E) (2002)}.
\bibitem{LT2} J. P. Ralston and B. Pire, 
\href{https://doi.org/10.1103/PhysRevD.66.111501} {Phys. Rev. D {\bf 66}, 111501 (2002)}.
\bibitem{LT3} M. Burkardt, 
\href{https://doi.org/10.1142/S0217751X03012370} {Int. J. Mod. Phys. A {\bf 18}, 173 (2003)}.
\bibitem{LT4} W. Broniowski and E. Ruiz Arriola, 
\href{https://doi.org/10.1016/j.physletb.2003.09.009} {Phys. Lett. B {\bf 574}, 57 (2003)}.
\bibitem{Bakulev2000} A. P. Bakulev, R. Ruskov, K. Goeke and N. G. Stefanis,
\href{https://doi.org/10.1103/PhysRevD.62.054018} {Phys. Rev. D {\bf 62}, 054018 (2000)}.
\bibitem{Goeke2001} K. Goeke, M. V. Polyakov and M. Vanderhaeghen,
\href{https://doi.org/10.1016/S0146-6410(01)00158-2} {Prog. Part. Nucl. Phys. {\bf 47}, 401 (2001)}.
\bibitem{Diehl2003} M. Diehl,
\href{https://doi.org/10.1016/j.physrep.2003.08.002} {Phys. Rep. {\bf 388}, 41 (2003)}.
\bibitem{Belitsky2005} A. V. Belitsky and  A. V. Radyushkin,
\href{https://doi.org/10.1016/j.physrep.2005.06.002} {Phys. Rep. {\bf 418}, 1 (2005)}.
\bibitem{Boffi2007} S. Boffi and  B. Pasquini,
\href{https://doi.org/10.1393/ncr/i2007-10025-7} {Riv. Nuovo Cim. {\bf 30}, 387 (2007)}.
\bibitem{Ji1997} X. D. Ji, 
\href{https://doi.org/10.1103/PhysRevD.55.7114} {Phys. Rev. D {\bf 55}, 7114 (1997)}.
\bibitem{BEG2018} V. D. Burkert, L. Elouadrhiri and F. X. Girod 
\href{https://doi.org/10.1038/s41586-018-0060-z} {Nature (London) {\bf 557}, 396 (2018)}.
\bibitem{NJL1} W. Broniowski, A. E. Dorokhov and E. Ruiz Arriola, 
\href{https://doi.org/10.1103/PhysRevD.82.094001} {Phys. Rev. D {\bf 82}, 094001 (2010)}.
\bibitem{NJL2} A. E. Dorokhov, W. Broniowski  and E. Ruiz Arriola, 
\href{https://doi.org/10.1103/PhysRevD.84.074015} {Phys. Rev. D {\bf 84}, 074015 (2011)}.
\bibitem{NJL3} A. Freese and I. C. Clo{\"e}t, 
\href{https://doi.org/10.1103/PhysRevC.100.015201} {Phys. Rev. C {\bf 100}, 015201 (2019)}.[Erratum: \href{https://doi.org/10.1103/PhysRevC.105.059901} {Phys. Rev. C {\bf 105}, 059901 (2022)}]
\bibitem{LQCD1} D. Br{\"o}mmel et al. (QCDSF/UKQCD Collaborations), 
\href{https://doi.org/10.1103/PhysRevLett.101.122001} {Phys. Rev. Lett. {\bf 101}, 122001 (2008)}.
\bibitem{LQCD2} D. Br{\"o}mmel et al. (QCDSF/UKQCD Collaborations), 
\href{https://doi.org/10.1140/epjc/s10052-007-0295-6} {Eur. Phys. J. C {\bf 51}, 335 (2007)}.
\bibitem{PiGPD1} M. V. Polyakov and C. Weiss, 
\href{https://doi.org/10.1103/PhysRevD.60.114017} {Phys. Rev. D {\bf 60}, 114017 (1999)}.
\bibitem{PiGPD2} W. Broniowski, E. Ruiz Arriola and K. Golec-Biernat, 
\href{https://doi.org/10.1103/PhysRevD.77.034023} {Phys. Rev. D {\bf 77}, 034023 (2008)}.
\bibitem{PiGPD3} C. Mezrag et al., 
\href{https://doi.org/10.1016/j.physletb.2014.12.027} {Phys. Lett. B {\bf 741}, 190 (2015)}.
\bibitem{PiGPD4} C. Fanelli et al., 
\href{https://doi.org/10.1140/epjc/s10052-016-4101-1} {Eur. Phys. J. C {\bf 76}, 253 (2016)}.
\bibitem{PiGPD5} G. F. de Teramond et al. (HLFHS Collaboration), 
\href{https://doi.org/10.1103/PhysRevLett.120.182001} {Phys. Rev. Lett {\bf 120}, 182001 (2018)}.
\bibitem{PiGPD6} J. W. Chen, H. W. Lin and J. H. Zhang, 
\href{https://doi.org/10.1016/j.nuclphysb.2020.114940} {Nucl. Phys. B. {\bf 952}, 114940 (2020)}.
\bibitem{PiGPD7} C. Shi, K. Bednar, I. C. Clo{\"e}t and A. Freese, 
\href{https://doi.org/10.1103/PhysRevD.101.074014} {Phys. Rev. D {\bf 101}, 074014 (2020)}. 
\bibitem{PiGPD8} J. L. Zhang et al., 
\href{https://doi.org/10.1016/j.physletb.2021.136158} {Phys. Lett. B {\bf 815}, 136158 (2021)}.
\bibitem{PiGPD9} J. L. Zhang, Z. F. Cui, J, Ping and C. D. Roberts,
\href{https://doi.org/10.1140/epjc/s10052-020-08791-1} {Eur. Phys. J. C {\bf 81}, 6 (2021)}.
\bibitem{PiGPD10} C. D. Roberts, D. G. Richards, T. Horn and L. Chang, 
\href{https://doi.org/10.1016/j.ppnp.2021.103883} {Prog. Part. Nucl. Phys. {\bf 120}, 103883 (2021)}.
\bibitem{PiGPD11} L. Adhikari et al. (BLFQ Collaboration),
\href{https://doi.org/10.1103/PhysRevD.104.114019} {Phys. Rev. D {\bf 104}, 114019 (2021)}.
\bibitem{PiGPD12} J. M. M. Chavez et al.,
\href{https://doi.org/10.1103/PhysRevD.105.094012} {Phys. Rev. D {\bf 105}, 094012 (2022)}.
\bibitem{PiGPD13} K. Raya et al.,
\href{https://doi.org/10.1088/1674-1137/ac3071} {Chin. Phys. C {\bf 46}, 013105 (2022)}.
\bibitem{PiGPD14} Y. Guo, X. Ji and K. Shiells,
\href{https://doi.org/10.1007/JHEP09(2022)215} {JHEP {\bf 09}, 215 (2022)}.
\bibitem{PiChi1} W. Broniowski and E. Ruiz Arriola,
\href{https://doi.org/10.1103/PhysRevD.78.094011} {Phys. Rev. D {\bf 78}, 094011 (2008)}.
\bibitem{PiChi2} H. D. Son and H. C. Kim,
\href{https://doi.org/10.1103/PhysRevD.90.111901} {Phys. Rev. D {\bf 90}, 111901 (2014)}.
\bibitem{PiFF1} W. Broniowski, E. Ruiz Arriola and P. Sanchez-Puertas,
\href{https://doi.org/10.1103/PhysRevD.106.036001} {Phys. Rev. D {\bf 106}, 036001 (2022)}.
\bibitem{PiFF2} A. F. Krutov and V. E. Troitsky,
\href{https://doi.org/10.1103/PhysRevD.103.014029} {Phys. Rev. D {\bf 103}, 014029 (2021)}.
\bibitem{PiFF3} Z. Xing, M. Ding and L. Chang,
\href{https://doi.org/10.1103/PhysRevD.107.L031502} {Phys. Rev. D {\bf 107}, L031502 (2023)}.
\bibitem{PiFF4} Y. Z. Xu et al.,
\href{https://doi.org/10.1140/epjc/s10052-024-12518-x} {Eur. Phys. J. C {\bf 84}, 191 (2024)}.
\bibitem{Raya2024} I. M. Higuera-Angulo, R. J. Hernandez-pinto, K. Raya and A. Bashir,
\href{https://doi.org/10.1103/PhysRevD.110.034013}{Phys. Rev. D {\bf 110} 034013 (2024)}.
\bibitem{Raya106} L. Albino, I. M. Higuera-Angulo, K. Raya and A. Bashir,
\href{https://doi.org/10.1103/PhysRevD.106.034003} {Phys. Rev. D {\bf 106}, 034003 (2022)}.
\bibitem{Raya2025} K. Raya, A. Bashir and J. Rodríguez-Quintero,
\href{https://doi.org/10.1088/0256-307X/42/2/020201} {Chin. Phys. Lett. {\bf 42}, 020201 (2025)}.
\bibitem{Diehl2001} M. Diehl, T. Feldmann, R. Jakob and P.Kroll,
\href{https://doi.org/10.1016/S0550-3213(00)00684-2;https://doi.org/10.1016/S0550-3213(01)00183-3} {Nucl. Phys. B {\bf 596}, 33 (2001); {\bf 605}, 647 (E) (2001)}.
\bibitem{LFWFs1} G. R. Goldstein, J. O. Hernandez and S. Liuti,
\href{https://doi.org/10.1103/PhysRevD.84.034007} {Phys. Rev. D {\bf 84}, 034007 (2011)}.
\bibitem{LFWFs2} J. O. Hernandez, S. Liuti, G. R. Goldstein and K. Kthuria,
\href{https://doi.org/10.1103/PhysRevC.88.065206} {Phys. Rev. C {\bf 88}, 065206 (2013)}.
\bibitem{LFWFs3} N. Kumar, C. Mondal and N. Sharma,
\href{https://doi.org/10.1140/epja/i2017-12433-0} {Eur. Phys. J. A {\bf 53}, 237 (2017)}.
\bibitem{LFWFs4} B. Kriesten,
\href{https://doi.org/10.1103/PhysRevD.105.056022} {Phys. Rev. D {\bf 105}, 056022 (2022)}.
\bibitem{LFWFs5} Y. Liu et al. (BLFQ Collaboration),
\href{https://doi.org/10.1103/PhysRevD.105.094018} {Phys. Rev. D {\bf 105}, 094018 (2022)}.
\bibitem{CMMR} N. Chouika, C. Mezrag, H. Moutarde and J. Rodríguez-Quintero,
\href{https://doi.org/10.1016/j.physletb.2018.02.070} {Eur. Phys. J. C {\bf 77}, 906 (2017)}.
\bibitem{CMMR1} N. Chouika, C. Mezrag, H. Moutarde and J. Rodríguez-Quintero,
\href{https://doi.org/10.1140/epjc/s10052-017-5465-6} {Phys. Lett. B {\bf 780}, 287 (2018)}.
\bibitem{NA2025} H.Nematollahi and K. Azizi,
\href{https://doi.org/10.1103/PhysRevD.111.014011} {Phys. Rev. D {\bf 111}, 014011 (2025)}.
\bibitem{GA22} H. Hashamipour, M. Goharipour, K. Azizi and V. Goloskokov,
\href{https://doi.org/10.1103/PhysRevD.105.054002} {Phys. Rev. D {\bf 105}, 054002 (2022)}.
\bibitem{GA231} H. Hashamipour, M. Goharipour, K. Azizi and V. Goloskokov,
\href{https://doi.org/10.1103/PhysRevD.107.096005} {Phys. Rev. D {\bf 107}, 096005 (2023)}.
\bibitem{GA232} F. Irani, M. Goharipour, H. Hashamipour and K. Azizi,
\href{https://doi.org/10.1103/PhysRevD.108.074018} {Phys. Rev. D {\bf 108}, 074018 (2023)}.
\bibitem{GA24} M. Goharipour, H. Hashamipour, F. Irani and K. Azizi,
\href{https://doi.org/10.1103/PhysRevD.109.074042} {Phys. Rev. D {\bf 109}, 074042 (2024)}.
 \bibitem{NYVPi} H. Nematollahi and M. M. Yazdanpanah,
\href{https://doi.org/10.1016/j.nuclphysa.2018.05.009} {Nucl. Phys. A {\bf 977}, 23 (2018)}.
\bibitem{Chi5} A. Watanabe, C. W. Kao and K. Suzuki,
\href{https://doi.org/10.1103/PhysRevD.94.114008} {Phys. Rev. D {\bf 94}, 114008 (2016)}.
\bibitem{Chi6} A. Watanabe, T. Sawada and C. W. Kao,
\href{https://doi.org/10.1103/PhysRevD.97.074015} {Phys. Rev. D {\bf 97}, 074015 (2018)}.
\bibitem{NYSGPi} H. Nematollahi and M. M. Yazdanpanah,
\href{https://doi.org/0.1140/epjp/i2019-12844-2} {Eur. Phys. J. Plus {\bf 134}, 382 (2019)}.
\bibitem{Chi1} K. Suzuki and W. Weise,
\href{https://doi.org/10.1016/S0375-9474(98)00132-8} {Nucl. Phys. A {\bf 634}, 141 (1998)}.
\bibitem{Chi2} Y. Ding, R-G. Xu and B-Q. Ma,
\href{https://doi.org/10.1103/PhysRevD.71.094014} {Phys. Rev. D {\bf 71}, 094014 (2005)}.
\bibitem{Chi3} H. Song, X. Zhang and B-Q. Ma,
\href{https://doi.org/10.1140/epjc/s10052-011-1542-4} {Eur. Phys. J. C {\bf 71}, 1542 (2011)}. 
\bibitem{Chi4} H. Nematollahi, M. M. Yazdanpanah and A. Mirjalili,
\href{https://doi.org/10.1088/0954-3899/39/4/045009} {J. Phys. G: Nucl. Part. Phys. {\bf 39}, 045009 (2012)}.
\bibitem{AP} G. Altarelli and G. Parisi,
\href{https://doi.org/10.1016/0550-3213(77)90384-4} {Nucl. Phys. B {\bf 126}, 298 (1977)}.
\bibitem{HCT} P.T.P. Hutauruk, I. C. Cloet and A. W. Thomas,
\href{https://doi.org/10.1103/PhysRevC.94.035201} {Phys. Rev. C {\bf 94}, 035201 (2016)}.
\bibitem{Chenetal} C. Chen et al.,
\href{https://doi.org/10.1103/PhysRevD.93.074021} {Phys. Rev. D {\bf 93}, 074021 (2016)}.
\bibitem{Brodsky1989} S. J. Brodsky and G. P. Lepage,
\href{https://doi.org/10.1142/9789814503266_0002} {Adv. Ser. Direct. High Energy Phys. {\bf 5}, 93 (1989)}.
\bibitem{Alex2022} C. Alexandrou et al.,
\href{https://doi.org/10.1103/PhysRevD.105.054502} {Phys. Rev. D {\bf 105}, 054502 (2022)}.
\bibitem{ErA2022} N. Er and K. Azizi,
\href{https://doi.org/10.1140/epjc/s10052-022-10333-w} {Eur. Phys. J. C {\bf 82}, 397 (2022)}.
\bibitem{Dally1980} E. B. Dally et al.,
\href{https://doi.org/10.1103/PhysRevLett.45.232} {Phys. Rev. Lett. {\bf 45}, 232 (1980)}.
\bibitem{Amend1986} S. R. Amendolia et al.,
\href{https://doi.org/10.1016/0370-2693(86)91407-3} {Phys. Lett. B {\bf 178}, 435 (1986)}.
\bibitem{JLAB1} T. Horn and C. D. Roberts,
\href{https://doi.org/10.1088/0954-3899/43/7/073001} {J. Phys. G {\bf 43}, 073001 (2016)}.
\bibitem{JLAB2} M. Coman et al. (Jefferson Lab Hall A Collaboration),
\href{https://doi.org/10.1103/PhysRevC.81.052201} {Phys. Rev. C {\bf 81}, 052201 (2010)}.
\bibitem{Gao2017} F. Chao, L. Chang, Y. X. Liu, C. D. Roberts and P. C. Tandy,
\href{https://doi.org/10.1103/PhysRevD.96.034024} {Phys. Rev. D {\bf 96}, 034024 (2017)}.
\bibitem{PiRad1} P. Zyla et al.,
\href{https://doi.org/10.1093/ptep/ptaa104} {PTEP {\bf 2020}, 083C01 (2020)}.
\bibitem{PiRad2} Z. F. Cui et al.,
\href{https://doi.org/10.1016/j.physletb.2021.136631} {Phys. Lett. B {\bf 822}, 136631 (2021)}.

\end{thebibliography}
\end{document}